\documentstyle[12pt,qqa4lart]{article}
\input tcilatex

\QQQ{Language}{
American English
}

\begin{document}

\author{Lu-Ming Duan and Guang-Can Guo\thanks{%
Electronic address: gcguo@sunlx06.nsc.ustc.edu.cn} \\
Department of Physics and Nonlinear Science Center,\\
University of Science and Technology of China, \\
Hefei, Anhui 230026, People's Republic of China}
\title{Reducing decoherence in quantum computer memory with all quantum bits
coupling to the same environment}
\date{}
\maketitle

\begin{abstract}
\baselineskip 24pt Decoherence in quantum computer memory due to the
inevitable coupling to the external environment is examined. We take the
assumption that all quantum bits (qubits) interact with the same environment
rather than the assumption of separate environments for different qubits. It
is found that the qubits are decohered collectively. For some kinds of
entangled input states, no decoherence occurs at all in the memory even if
the qubits are interacting with the environment. Based on this phenomenon, a
scheme is proposed for reducing the collective decoherence. We also discuss
possible implications of this decoherence model for quantum measurements.\\

\end{abstract}

\newpage\baselineskip 24pt

Quantum computers have recently raised a lot of interest [1,2]. In quantum
computers, the contents of the memory cells are in a superposition of
different states and the computer performs deterministic unitary
transformations on the quantum states of these memory cells [3]. A two-state
memory cell, which may be a spin-$\frac 12$ electron or a two-level atom, is
called a quantum bit, or qubit [4]. It has been argued that quantum
computers can solve certain problems much more efficiently than the
classical computers [5-8]. An impressive example is, Shor [8] has shown a
quantum computer could solve the problem of finding factors of a large
number $N$ in a time which is a polynomial function of the length $L$
(number of bits) of the number. However, it is not yet clear whether quantum
computers are feasible to build. Decoherence is one of the major obstacles
to realizing quantum computation. It has been found that decoherence in
quantum computer memory can not be neglected if the qubits interact with the
external environment [9,10]. To reduce this decoherence, some strategies,
such as the quantum error-correction [11-17] and the purification of noisy
entanglement [18,19], have been proposed.

In the previous analysis of decoherence [9,10], an important assumption has
been taken. This is that the qubits are assumed to couple independently to
separate environments. Independent decoherence is an ideal case. As has been
pointed out by Ekert and Lloyd [9], there is another practical circumstance,
in which the qubits interact with the same environment. The interaction with
the same environment will result in cooperative decoherence for the qubits.
Plama et al. [20] provided the first step in studying the cooperative
decoherence. They started by considering a system of two qubits and extended
the result to include a register of $L$ qubits. In this paper, we propose an
alternative simple approach for studying the collective decoherence and
correct an error in the calculation of Ref. [20]. We consider a system of $L$
qubits interacting with the same environment. It is found that the qubits
are decohered collectively. This is compared with the independent
decoherence of the qubits when they interact with separate environments.
Because of the collective decoherence, for some kinds of input states
(called the coherence-preserving states), no decoherence occurs at all even
if the qubits are interacting with the environment. Based on this
phenomenon, a simple scheme is proposed for reducing the collective
decoherence. The coherence-preserving states are intimately related to the
concept of the point basis introduced in the theory of quantum measurements
[21]. We also discuss possible implications of this decoherence model for
quantum measurements.

Now we consider the decoherence model---$L$ qubits in the memory jointly
coupled to the same environment. The $l$ qubit can be described by the Pauli
operator $\overrightarrow{\sigma _l}$. The environment is modelled by a bath
of oscillators. We consider the decoherence resulting from the dephasing
process. The Hamiltonian describing the phase damping has the form [22] 
\begin{equation}
\label{1}H=\hbar \left\{ \int d\omega \left[ \stackunder{l=1}{\stackrel{L}{%
\sum }}\kappa \left( \omega \right) \left( a_\omega +a_\omega ^{+}\right)
\sigma _l^z+\omega a_\omega ^{+}a_\omega \right] \right\} ,
\end{equation}
where $L$ indicates number of the qubits. We have supposed that the coupling
constants $\kappa \left( \omega \right) $ to the environment are the same
for all the qubits. In the Hamiltonian (1), $\sigma _l^z$ $\left(
l=1,2,\cdots ,L\right) $ are conservative operators, so the dynamical
equations for the operators can be easily solved. The Heisenberg equation
for the bath operator is 
\begin{equation}
\label{2}i\stackrel{.}{a}_\omega =\omega a_\omega +\stackunder{l=1}{%
\stackrel{L}{\sum }}\kappa \left( \omega \right) \sigma _l^z.
\end{equation}
It has the solution 
\begin{equation}
\label{3}a_\omega \left( t\right) =a_\omega \left( 0\right) e^{-i\omega t}+%
\stackunder{l=1}{\stackrel{L}{\sum }}\kappa \left( \omega \right) \sigma _l^z%
\frac{e^{-i\omega t}-1}\omega .
\end{equation}

To determine the magnitude of decoherence, we need know the evolution of the
reduced density of the qubits. This problem can be solved by using the
operator representation of the density operator [23]. Let $\rho
_{-1,-1}=\frac 12\left( I+\sigma ^z\right) ,$ $\rho _{1,1}=\frac 12\left(
I-\sigma ^z\right) ,$ $\rho _{-1,1}=\sigma ^{+},$ $\rho _{1,-1}=\sigma ^{-}$%
, and 
\begin{equation}
\label{4}\rho _{\left\{ i_l,j_l\right\} }=\rho _{i_1,j_1}\otimes \rho
_{i_2,j_2}\otimes \cdots \otimes \rho _{i_L,j_L},
\end{equation}
where $I$ is the unit operator, and possible values for $i_l$ and $j_l$ $%
\left( l=1,2,\cdots ,L\right) $ are $1$ or $-1$. All the operators defined
by Eq. (4) are expressed by Pauli's operators. Obviously, they make a
complete set. The initial density operator of the qubits can be expanded
into the set of operators $\rho _{\left\{ i_l,j_l\right\} }$. Suppose the
environment is initially in thermal equilibrium. The total density operator $%
\rho \left( 0\right) $ is then expressed as 
\begin{equation}
\label{5}
\begin{array}{c}
\rho \left( 0\right) =\rho _s\left( 0\right) \otimes \rho _{env}\left(
0\right)  \\  
\\ 
=\stackunder{\left\{ i_l,j_l\right\} }{\sum }c_{\left\{ i_l,j_l\right\}
}\rho _{\left\{ i_l,j_l\right\} }\left( 0\right) \otimes \stackunder{\omega 
}{\prod }\int d^2\alpha _\omega \frac 1{\pi \left\langle N_\omega
\right\rangle }\exp \left( -\frac{\left| \alpha _\omega \right| ^2}{%
\left\langle N_\omega \right\rangle }\right) \left( \left| \alpha _\omega
\right\rangle \left\langle \alpha _\omega \right| \right) _0,
\end{array}
\end{equation}
where the subscripts $s$ and $env$ denote the system (qubits) and the
environment, respectively. $\left\langle N_\omega \right\rangle $ is the
mean photon number of the bath mode $\omega $ 
\begin{equation}
\label{6}\left\langle N_\omega \right\rangle =\frac 1{e^{\hbar \omega
/k_BT}-1}.
\end{equation}

In the Schr\"odinger picture, the density operator obeys Von Neumann's
equation which differs from Heisenberg's equation by an overall sign.
Therefore, the density operator (in the Schr\"odinger picture) may be
treated as an ordinary operator (in the Heisenberg picture) evolving
backwards in time. We thus have 
\begin{equation}
\label{7}\rho \left( t\right) =\stackunder{\left\{ i_l,j_l\right\} }{\sum }%
c_{\left\{ i_l,j_l\right\} }\rho _{\left\{ i_l,j_l\right\} }\left( -t\right)
\otimes \stackunder{\omega }{\prod }\int d^2\alpha _\omega \frac 1{\pi
\left\langle N_\omega \right\rangle }\exp \left( -\frac{\left| \alpha
_\omega \right| ^2}{\left\langle N_\omega \right\rangle }\right) \left(
\left| \alpha _\omega \right\rangle \left\langle \alpha _\omega \right|
\right) _{-t}, 
\end{equation}
where $\rho _{\left\{ i_l,j_l\right\} }$ and $\left| \alpha _\omega
\right\rangle \left\langle \alpha _\omega \right| $ are treated as ordinary
operators. From Eq. (3), we know 
\begin{equation}
\label{8}\left( \left| \alpha _\omega \right\rangle \left\langle \alpha
_\omega \right| \right) _{-t}=\left| \alpha _\omega e^{-i\omega t}+%
\stackunder{l=1}{\stackrel{L}{\sum }}\kappa \left( \omega \right) \sigma _l^z%
\frac{e^{-i\omega t}-1}\omega \right\rangle \left\langle \alpha _\omega
e^{-i\omega t}+\stackunder{l=1}{\stackrel{L}{\sum }}\kappa \left( \omega
\right) \sigma _l^z\frac{e^{-i\omega t}-1}\omega \right| _0. 
\end{equation}
So the reduced density operator of the qubits at time $t$ can be expressed
as 
\begin{equation}
\label{9}\rho _s\left( t\right) =Tr_{env}\left( \rho \left( t\right) \right)
=\stackunder{\left\{ i_l,j_l\right\} }{\sum }c_{\left\{ i_l,j_l\right\}
}\otimes \int \left\langle \{\alpha _\omega \}\right| \rho _{\left\{
i_l,j_l\right\} }\left( -t\right) \left| \{\alpha _\omega \}\right\rangle 
\stackunder{\omega }{\prod }\left[ P_\omega (\alpha _\omega ,t)d^2\alpha
_\omega \right] , 
\end{equation}
where $\left| \{\alpha _\omega \}\right\rangle =\stackunder{\omega }{\prod }%
\otimes \left| \alpha _\omega \right\rangle $ indicates the co-eigenstates
of all the bath operators $a_\omega \left( 0\right) $ and 
\begin{equation}
\label{10}P_\omega (\alpha _\omega ,t)=\frac 1{\pi \left\langle N_\omega
\right\rangle }\exp \left[ -\frac 1{\left\langle N_\omega \right\rangle
}\left| \alpha _\omega e^{i\omega t}+\stackunder{l=1}{\stackrel{L}{\sum }}%
\kappa \left( \omega \right) \sigma _l^z\frac{e^{i\omega t}-1}\omega \right|
^2\right] . 
\end{equation}

Now we need to obtain the diagonal matrix elements of the operator $\rho
_{\left\{ i_l,j_l\right\} }\left( -t\right) $ in the bath coherent
representation. From the Hamiltonian (1), the Heisenberg equation for the
operator $\rho _{\left\{ i_l,j_l\right\} }$ is 
\begin{equation}
\label{11}
\begin{array}{c}
i 
\stackrel{.}{\rho }_{\left\{ i_l,j_l\right\} }=\left[ \stackunder{l=1}{%
\stackrel{L}{\sum }}\left( i_l-j_l\right) \right] \left[ \int d\omega \kappa
\left( \omega \right) \left( a_\omega +a_\omega ^{+}\right) \right] \rho
_{\left\{ i_l,j_l\right\} } \\  \\ 
=\left[ 
\stackunder{l=1}{\stackrel{L}{\sum }}\left( i_l-j_l\right) \right] \left[
\int d\omega \kappa \left( \omega \right) \left( \rho _{\left\{
i_l,j_l\right\} }a_\omega \left( 0\right) e^{-i\omega t}+a_\omega ^{+}\left(
0\right) e^{i\omega t}\rho _{\left\{ i_l,j_l\right\} }\right) \right] \\  \\ 
-\int d\omega \kappa ^2\left( \omega \right) \left\{ \frac{\cos \left(
\omega t\right) -1}\omega \left[ \left( \stackunder{l=1}{\stackrel{L}{\sum }}%
i_l\right) ^2-\left( \stackunder{l=1}{\stackrel{L}{\sum }}j_l\right)
^2\right] +i\frac{\sin \left( \omega t\right) }\omega \left[ \stackunder{l=1%
}{\stackrel{L}{\sum }}\left( i_l-j_l\right) \right] ^2\right\} \rho
_{\left\{ i_l,j_l\right\} }. 
\end{array}
\end{equation}
In the derivation, the following relations are used: 
\begin{equation}
\label{12}\rho _{\left\{ i_l,j_l\right\} }\sigma _l^z=-j_l\rho _{\left\{
i_l,j_l\right\} }, 
\end{equation}
\begin{equation}
\label{13}\sigma _l^z\rho _{\left\{ i_l,j_l\right\} }=-i_l\rho _{\left\{
i_l,j_l\right\} }. 
\end{equation}
The solution of Eq. (11) is 
\begin{equation}
\label{14}
\begin{array}{c}
\rho _{\left\{ i_l,j_l\right\} }\left( t\right) =\exp \left\{ i\int d\omega
\kappa ^2\left( \omega \right) \left\{ 
\frac{\sin \left( \omega t\right) -\omega t}{\omega ^2}\left[ \left( 
\stackunder{l=1}{\stackrel{L}{\sum }}i_l\right) ^2-\left( \stackunder{l=1}{%
\stackrel{L}{\sum }}j_l\right) ^2\right] \right. \right. \\  \\ 
\left. \left. +i 
\frac{1-\cos \left( \omega t\right) }{\omega ^2}\left[ \stackunder{l=1}{%
\stackrel{L}{\sum }}\left( i_l-j_l\right) \right] ^2\right\} \right\} \\  \\ 
\times \exp \left\{ -\left[ 
\stackunder{l=1}{\stackrel{L}{\sum }}\left( i_l-j_l\right) \right] \int
d\omega \kappa \left( \omega \right) \frac{e^{i\omega t}-1}\omega a_\omega
^{+}\left( 0\right) \right\} \rho _{\left\{ i_l,j_l\right\} }\left( 0\right)
\\  \\ 
\times \exp \left\{ \left[ \stackunder{l=1}{\stackrel{L}{\sum }}\left(
i_l-j_l\right) \right] \int d\omega \kappa \left( \omega \right) \frac{%
e^{-i\omega t}-1}\omega a_\omega \left( 0\right) \right\} . 
\end{array}
\end{equation}

The operator $\rho _{\left\{ i_l,j_l\right\} }\left( t\right) $, so $\rho
_{\left\{ i_l,j_l\right\} }\left( -t\right) $, has been expressed by $\rho
_{\left\{ i_l,j_l\right\} }\left( 0\right) $ and $a_\omega \left( 0\right) $%
. Therefore, $\left\langle \{\alpha _\omega \}\right| \rho _{\left\{
i_l,j_l\right\} }\left( -t\right) \left| \{\alpha _\omega \}\right\rangle $
is obtained as a function of $\rho _{\left\{ i_l,j_l\right\} }\left(
0\right) $ and $\alpha _\omega $. Substituting this result into Eq. (9), we
thus have 
\begin{equation}
\label{15}
\begin{array}{c}
\rho _s\left( t\right) = 
\stackunder{\left\{ i_l,j_l\right\} }{\sum }c_{\left\{ i_l,j_l\right\} }\exp
\left\{ -\eta \left( t\right) \left[ \stackunder{l=1}{\stackrel{L}{\sum }}%
\left( i_l-j_l\right) \right] ^2\right\} \\  \\ 
\cdot \exp \left\{ i\Delta \phi \left( t\right) \left[ \left( \stackunder{l=1%
}{\stackrel{L}{\sum }}i_l\right) ^2-\left( \stackunder{l=1}{\stackrel{L}{%
\sum }}j_l\right) ^2\right] \right\} \cdot \rho _{\left\{ i_l,j_l\right\}
}\left( 0\right) , 
\end{array}
\end{equation}
where the Lamb phase shift factor $\Delta \phi \left( t\right) $ is defined
as 
\begin{equation}
\label{16}\Delta \phi \left( t\right) =\int d\omega \kappa ^2\left( \omega
\right) \left[ \frac{\omega t-\sin \left( \omega t\right) }{\omega ^2}%
\right] 
\end{equation}
and the phase damping factor $\eta \left( t\right) $ is 
\begin{equation}
\label{17}\eta \left( t\right) =\int d\omega \kappa ^2\left( \omega \right) 
\frac{4\sin ^2\left( \frac{\omega t}2\right) }{\omega ^2}\left( \left\langle
N_\omega \right\rangle +\frac 12\right) . 
\end{equation}
Both the Lamb phase shift and the phase damping have contributions to
decoherence of the qubits. The Lamb phase shift was missed in Ref. [20]. By
examining the calculation there, we find there is a mistake. In Eq. (13) of
Ref. [20], the time evolution operator $U(t)$ in the interaction picture is
expressed as $U\left( t\right) =\exp \left[ -\frac i\hbar \int_0^tH_I\left(
t^{^{\prime }}\right) dt^{^{\prime }}\right] $. But this expression is not
correct since there $\left[ H_I\left( t\right) ,H_I\left( t^{^{\prime
}}\right) \right] \neq 0$. Because of this error, a phase factor was missed
in the evolution operator $U(t)$. This phase factor finally results in the
Lamb phase shift. From Eq. (15), we see, in the case of collective
decoherence, the Lamb phase shift does not reduce to zero. Eq. (17) shows
that the phase damping is directly proportional to the mean photon number.
At high temperature, i.e., $k_BT>>\hbar \omega $, decoherence is mainly
induced by the phase damping. But at low temperature, the Lamb phase shift
is of the same order of magnitude as the phase damping and it can not be
neglected.

The state fidelity has been introduced to describe stability of quantum
information [24]. For a pure input state $\left| \Psi \left( 0\right)
\right\rangle $, the fidelity is defined as 
\begin{equation}
\label{18}F=\left\langle \Psi \left( 0\right) \right| \rho _s\left( t\right)
\left| \Psi \left( 0\right) \right\rangle =tr\left[ \rho _s\left( 0\right)
\rho _s\left( t\right) \right] ,
\end{equation}
where $\rho _s\left( 0\right) =\left| \Psi \left( 0\right) \right\rangle
\left\langle \Psi \left( 0\right) \right| $ and $\rho _s\left( t\right) $
indicates the output density operator of the system. Here we use the
fidelity to describe decoherence of the qubits. Suppose the initial state of
the qubits is expressed as $\left| \Psi \left( 0\right) \right\rangle =%
\stackunder{\left\{ i_l\right\} }{\sum }c_{\left\{ i_l\right\} }\left|
i_l\right\rangle $, where $\left| i_l\right\rangle $ with $i_l=\pm 1$ may
represent the states $\left| \pm \frac 12\right\rangle $ of a spin-$\frac 12$
electron, or the states $\left| e\right\rangle $ and $\left| g\right\rangle $
of a two-level atom. From Eq. (15), we get the fidelity 
\begin{equation}
\label{19}
\begin{array}{c}
F=
\stackunder{\left\{ i_l,j_l\right\} }{\sum }\left| c_{\left\{ i_l\right\}
}\right| ^2\left| c_{\left\{ j_l\right\} }\right| ^2\exp \left\{ -\eta
\left( t\right) \left[ \stackunder{l=1}{\stackrel{L}{\sum }}\left(
i_l-j_l\right) \right] ^2\right\}  \\  \\ 
\cdot \exp \left\{ i\Delta \phi \left( t\right) \left[ \left( \stackunder{l=1%
}{\stackrel{L}{\sum }}i_l\right) ^2-\left( \stackunder{l=1}{\stackrel{L}{%
\sum }}j_l\right) ^2\right] \right\} 
\end{array}
\end{equation}
In the derivation, the following relations are used. 
\begin{equation}
\label{20}tr\left( \rho _{1,-1}\rho _{-1,1}\right) =tr\left( \rho
_{-1,-1}^2\right) =tr\left( \rho _{1,1}^2\right) =1,
\end{equation}
\begin{equation}
\label{21}
\begin{array}{c}
tr\left( \rho _{1,1}\rho _{1,-1}\right) =tr\left( \rho _{1,1}\rho
_{-1,1}\right) =tr\left( \rho _{-1,-1}\rho _{1,-1}\right)  \\  
\\ 
=tr\left( \rho _{-1,-1}\rho _{-1,1}\right) =tr\left( \rho _{1,1}\rho
_{-1,-1}\right) =0.
\end{array}
\end{equation}

Eqs. (15) and (19) suggest, the qubits are decohered collectively. This is
an interesting phenomenon. The phase shift and the phase damping are
directly proportional to the factor $\stackunder{l=1}{\stackrel{L}{\sum }}%
\left( i_l-j_l\right) $. If the input density operator satisfies $%
\stackunder{l=1}{\stackrel{L}{\sum }}\left( i_l-j_l\right) =0$, at any time
the fidelity $F=\left[ \stackunder{\left\{ i_l\right\} }{\sum }\left|
c_{\left\{ i_l\right\} }\right| ^2\right] ^2=1$ and the reduced density
operator of the qubits $\rho _s\left( t\right) =\stackunder{\left\{
i_l,j_l\right\} }{\sum }c_{\left\{ i_l,j_l\right\} }\rho _{\left\{
i_l,j_l\right\} }\left( 0\right) =\rho _s\left( 0\right) $. So no
decoherence occurs at all even if the qubits are interacting with the
environment. The states satisfying the condition $\stackunder{l=1}{\stackrel{%
L}{\sum }}\left( i_l-j_l\right) =0$ are called the coherence-preserving
states. Consider the following states 
\begin{equation}
\label{22}\left| \Psi _m\left( 0\right) \right\rangle =\stackunder{\left\{
i_l\right\} \in A_m}{\sum }c_{\left\{ i_l\right\} }\left| \left\{
i_l\right\} \right\rangle ,
\end{equation}
where $m$ is a definite number and $A_m$ denotes the set $\left\{ i_l\left| 
\stackunder{l=1}{\stackrel{L}{\sum }}i_l=m\right. \right\} $. Obviously, the
relation $\stackunder{l=1}{\stackrel{L}{\sum }}\left( i_l-j_l\right) =0$ is
satisfied for this kind of states. So all the states (22) are the
coherence-preserving states. In these states, the qubits are entangled with
each other.

It is interesting to compare the collective decoherence with the independent
decoherence. If the qubits couple independently to separate environments,
similar to the derivation of Eq. (15), it is not difficult to obtain that at
time $t$ the reduced density $\rho _s^{^{\prime }}\left( t\right) $ of the
qubits is expressed as 
\begin{equation}
\label{23}
\begin{array}{c}
\rho _s^{^{\prime }}\left( t\right) =
\stackunder{\left\{ i_l,j_l\right\} }{\sum }c_{\left\{ i_l,j_l\right\} }\exp
\left[ -\eta \left( t\right) \stackunder{l=1}{\stackrel{L}{\sum }}\left(
i_l-j_l\right) ^2\right] \cdot \exp \left[ i\Delta \phi \left( t\right) 
\stackunder{l=1}{\stackrel{L}{\sum }}\left( i_l^2-j_l^2\right) \right] \cdot
\rho _{\left\{ i_l,j_l\right\} }\left( 0\right)  \\  \\ 
=\stackunder{\left\{ i_l,j_l\right\} }{\sum }c_{\left\{ i_l,j_l\right\}
}\exp \left[ -\eta \left( t\right) \stackunder{l=1}{\stackrel{L}{\sum }}%
\left( i_l-j_l\right) ^2\right] \cdot \rho _{\left\{ i_l,j_l\right\} }\left(
0\right) .
\end{array}
\end{equation}
The Lamb phase shift reduces to zero for the independent decoherence since
for $i_l=\pm 1$ we always have $i_l^2-j_l^2=0$. Eq. (23) shows that the
phase damping increases with $L$ (number of the qubits) monotonically. In
general, $\rho _s^{^{\prime }}\left( t\right) $ rapidly deviates from $\rho
_s^{^{\prime }}\left( 0\right) $ if $L$ is large. This can be clearly seen
from the state fidelity (indicated by $F^{^{\prime }}$) 
\begin{equation}
\label{24}F^{^{\prime }}=\stackunder{\left\{ i_l,j_l\right\} }{\sum }\left|
c_{\left\{ i_l\right\} }\right| ^2\left| c_{\left\{ j_l\right\} }\right|
^2\exp \left[ -\eta \left( t\right) \stackunder{l=1}{\stackrel{L}{\sum }}%
\left( i_l-j_l\right) ^2\right] 
\end{equation}
The typical behavior of $F^{^{\prime }}$ in the form of Eq. (24), as has
been discussed in Ref. [9], is $F^{^{\prime }}\propto e^{-\alpha \left(
t\right) L}$, i.e., the fidelity decays with $L$ exponentially. Its damping
is insensitive to the type of the initial states. This is much different
from decoherence of the qubits coupling to the same environment. In the
latter case, with some input states decoherence of the qubits may increase
with $L$ more rapidly. But with some other states, i.e., the
coherence-preserving states, no decoherence occurs at all. Sensitivity to
the type of the input states in an important property of the collective
decoherence.

To reduce the independent decoherence, many quantum error correction schemes
have been proposed [11-17]. The schemes in Ref. [11-16] are devised to
correct single qubit errors. In practice, one need repeatedly use these
schemes to correct errors. For the independent decoherence, it can be shown
easily that if $N$ error corrections are performed within a time interval $%
[0,T)$, there is a remaining error probability of order $O\left( \left(
T/N\right) ^2\right) $ after each error correction event [11]. Thus the
accumulated error at time $T$ is of order $O\left( N\left( T/N\right)
^2\right) $. This error can be made arbitrarily small by choosing a
sufficient large $N$. However, this analysis does not hold for the
collective decoherence, since in the latter case occurrence of errors for
different qubits is correlated. In fact, the error correction schemes are
not very efficient for reducing the collective decoherence, since they do
not take into account the specific interaction properties between the qubits
and the environment. Fortunately, for reducing the collective decoherence,
there is a simple and more efficient scheme. This scheme essentially
exploits the coherence-preserving states. Before storing a state into the
memory, we transform it into a coherence-preserving state in the form of Eq.
(22). The transformed state undergoes no decoherence in the noisy memory and
afterwards, it can be transformed back into the original state. In this
scheme we should find a one-to-one map from arbitrary input states onto the
coherence-preserving states in a larger Hilbert space. Suppose there are $2L$
qubits. The Hilbert space spanned by the coherence-preserving states (22)
with $\left\{ i_l\right\} \in A_0$ is indicated by $S_0$. The dimension of $%
S_0$ is $\left( 
\begin{array}{c}
2L \\ 
L
\end{array}
\right) $. If all the states in the space $S_0$ are efficiently used in the
transformation, the efficiency $\eta _m$ of this scheme attains
\begin{equation}
\label{25}\eta _m=\frac 1{2L}\log _2\left( 
\begin{array}{c}
2L \\ 
L
\end{array}
\right) \approx 1-\frac 1{4L}\log _2\left( \pi L\right) .
\end{equation}
The approximation is taken under the condition $L>>1$. So the maximum
efficiency is near to $1$ if $L$ is large. Of course, to make use of all the
states in $S_0$, it requires an involved encoding. A simple encoding ,
though it is not the most efficient, is to use two qubits to encode one
qubit. As has been mentioned in Ref. [20], the encoding is
\begin{equation}
\label{26}
\begin{array}{c}
\left| +1\right\rangle \longrightarrow \left| +1,-1\right\rangle , \\ 
\left| -1\right\rangle \longrightarrow \left| -1,+1\right\rangle .
\end{array}
\end{equation}
This encoding makes use of a subset of the coherence-preserving states in $%
S_0$. The encoding (26) can be easily fulfilled by using the quantum
controlled-NOT gates [25].

In this paper, the coherence-preserving states are obtained with the
assumption that the qubits in the memory undergo no amplitude damping. If
the amplitude damping in not negligible, the states (22) will not remain
unchanged. However, in Ref. [25], we developed a general method to set up
the coherence-preserving states. By a strategy called the free Hamiltonian
elimination, the coherence-preserving states are found to exist both for the
phase damping and for the amplitude damping, though they are not of the same
form. Furthermore, we shown there that the coherence-preserving states could
be operated on with quantum gates. These results suggest, the transformation
to the coherence-preserving states is a useful and efficient scheme for
reducing the collective decoherence.

The Hamiltonian (1) also describes decoherence of a spin-$\frac 12$ chain.
The spin chain can be adopted as a model of apparatus in some case.
Therefore, the decoherence model in this paper may also have some
implications for quantum measurements. In fact, the coherence-preserving
states discussed above are intimately related to the concept of the point
basis introduced by Zurek some years ago [21]. It has been recognized that
decoherence plays an essential role in quantum measurements [26].
Decoherence is induced by the inevitable interaction between the apparatus
and the environment. This decoherence causes the off-diagonal terms of the
density operator to decay in the point basis of the apparatus and leads to
the wave-packet collapse. The point basis consists of the eigenvectors of
the operator which commute with the apparatus-environment interaction
Hamiltonian. The coherence-preserving states just have this property, so
they make a point basis. It is nice to see how fast the off-diagonal terms
of the density decay in the point basis. To show this, we need to analyze
the time behaviors of $\eta \left( t\right) $ and $\Delta \phi \left(
t\right) $. At high temperature, $\eta \left( t\right) $ is much more
important. For the one-dimensional spin chain, $\kappa ^2\left( \omega
\right) $ has the form of $\kappa ^2\left( \omega \right) =\varepsilon
^2\omega /\hbar $, where $\varepsilon $ is approximately a constant [9]. In
the high temperature limit, Eq. (17) gives
\begin{equation}
\label{27}\eta \left( t\right) \approx \frac{\pi \varepsilon ^2k_BT}{\hbar ^2%
}t.
\end{equation}
The decoherence time is thus $\frac{\hbar ^2}{\pi \varepsilon ^2k_BT}$.
Comparing it with the decoherence time for harmonic oscillators, we see, for
the spin chain, the decoherence time follows a similar dependence on various
parameters (such as the coupling constant and temperature of the
environment) as is the case for harmonic oscillators.\\

{\bf Acknowledgments}

This project was supported by the National Natural Science Foundation of
China.

\newpage\ 

\end{document}